\documentclass[conference]{IEEEtran}
\IEEEoverridecommandlockouts
\usepackage{cite}
\usepackage{amsmath,amssymb,amsfonts}
\usepackage{algorithmic}
\usepackage{graphicx}
\usepackage{textcomp}
\usepackage{xcolor}
\usepackage{subfigure}
\usepackage{amsmath}
\usepackage{flushend}

\def\BibTeX{{\rm B\kern-.05em{\sc i\kern-.025em b}\kern-.08em
    T\kern-.1667em\lower.7ex\hbox{E}\kern-.125emX}}

\def\JUST{\text{JUNO}}

\begin{document}

\title{JUNO: Jump-Start Reinforcement Learning-based Node Selection for UWB Indoor Localization\\
\thanks{This Project was partially supported by Department of National Defence's Innovation for Defence Excellence \& Security (IDEaS), Canada.}
}

\author{\IEEEauthorblockN{1\textsuperscript{sh} Zohreh Hajiakhondi-Meybodi}
\IEEEauthorblockA{\textit{Electrical \& Computer Engineering,} \\
\textit{Concordia University,}\\
Montreal, Canada\\
z\_hajiak@encs.concordia.ca}
\and
\IEEEauthorblockN{2\textsuperscript{nd} Ming Hou}
\IEEEauthorblockA{\textit{Defence Research and Development} \\
\textit{Canada (DRDC),}\\
Toronto, Canada\\
ming.Hou@drdc-rddc.gc.ca}
\and
\IEEEauthorblockN{3\textsuperscript{th} Arash Mohammadi}
\IEEEauthorblockA{\textit{Concordia Institute for Inf. Systems Eng.,} \\
\textit{Concordia University,}\\
Montreal, Canada\\
arash.mohammadi@concordia.ca}
}

\maketitle

\begin{abstract}
Ultra-Wideband (UWB) is one of the key technologies empowering the  Internet of Thing (IoT) concept to perform reliable, energy-efficient, and highly accurate monitoring, screening, and localization in indoor environments. Performance of UWB-based localization systems, however, can significantly degrade because of Non Line of Sight (NLoS) connections between a mobile user and UWB beacons. To mitigate the destructive effects of NLoS connections, we target development of a Reinforcement Learning (RL) anchor selection framework that can efficiently cope with the dynamic nature of indoor environments. Existing RL models in this context, however, lack the ability to generalize well to be used in a new setting. Moreover, it takes a long time for the conventional RL models to reach the optimal policy. To tackle these challenges, we propose the Jump-start RL-based Uwb NOde selection (JUNO) framework, which performs real-time location predictions without relying on complex NLoS identification/mitigation methods. The effectiveness of the proposed $\JUST$  framework is evaluated in term of the location error, where the mobile user moves randomly through an ultra-dense indoor environment with a high chance of establishing NLoS connections. Simulation results corroborate the effectiveness of the proposed framework in comparison to its state-of-the-art counterparts.
\end{abstract}
\begin{IEEEkeywords}
Indoor Localization, Internet of Things, Anchor Selection, Jump-Start Reinforcement Learning, Ultra-Wideband (UWB).
\end{IEEEkeywords}
\section{Introduction} \label{sec:introduction}

Ultra-WideBand (UWB) technology has been emerged as a solution to meet the phenomenal growth of the need for localizing users in indoor environments~\cite{Huang2020}. The use of a wide radio spectrum in UWB technologies enables individual multi-path components of the received signal to be efficiently resolved, resulting in high accuracy indoor positioning~\cite{Zafari2019}. To monitor/track users in indoor environments, several localization techniques have been proposed such as Received Signal Strength Indicator (RSSI)~\cite{Atashi2020, Beni2020}, Angle of Arrival (AoA)~\cite{Hajiakhondi2020_1,Hajiakhondi2020_2,Hajiakhondi2021}, and Time Difference of Arrival (TDoA)~\cite{Khalaf-Allah2020,Huang2021}, among which time-based solutions~\cite{Khalaf-Allah2020,Huang2021,Zhao2021} can be considered as more efficient ones. In such scenarios, the time of the received signal from a set of available UWB beacons is required, which can only be estimated accurately if the first arrival path has been properly identified. In this context, one key challenge is susceptibility to the Non-Line-of-Sight (NLoS) error. In presence of an obstacle between the UWB beacon and the mobile device, the time of the received signal will be delayed, resulting in a positive bias and a significant degradation in the positioning accuracy. Therefore, NLoS mitigation/identification in UWB-based indoor localization is of paramount importance. Conventional indoor localization frameworks reduced such location error via parametric solutions~\cite{Zhang:2020, Zhang:2013}, the accuracy of which is dependent on implementation of complex pre-processing techniques adding considerable latency. Furthermore, using a large number of UWB beacons for localizing users is inefficient from the energy consumption perspective. In this regard, Dai~\textit{et al.}~\cite{Dai:2013} analytically proved that tracking users' locations through a subset of active beacons offers several benefits, including mitigating the energy consumption of beacons, and improving the location accuracy. Consequently, the main focus of recent researches~\cite{Wang:2020,  Albaidhani:2019, Courtay:2019, Albaidhani:2020} has been shifted to use anchor node selection to achieve the best localization performance in terms of location accuracy and resource management. The paper aims to further advance this emerging field.

\vspace{.025in}
\noindent
\textbf{Related Work:} Anchor node selection in the context of indoor localization is utilized to improve the network's performance by setting a set of criteria for selecting a subset of beacons with the highest utilities. One of the most important criteria in indoor localization is to mitigate the location error, caused by NLoS connections. Towards this goal, LoS connections will be selected for monitoring/tracking users' locations instead of extracting location information from all available beacons. Generally speaking, anchor selection frameworks can be classified into two groups, i.e., analytical solutions and Artificial Intelligence (AI) models. Conventionally, the focus of anchor selection frameworks was on the former category, i.e., deriving fixed mathematical/optimization models~\cite{Wang:2016, Albaidhani:2019, Albaidhani:2020} to meet the acceptable location accuracy.To alleviate the high complexity caused by using complicated mathematical formulations, recent research works~\cite{Nguyen:2020, Poulose:2020, Fan:2019, Salimibeni:2021} used AI and Machine Learning (ML) models to localize/navigate mobile devices in indoor environments. One of the most commonly used AI-based Los/NLoS identification approaches is supervised learning models~\cite{Nguyen:2020, Poulose:2020}. For instance, Poulose \textit{et al}.~\cite{Poulose:2020} applied Long Short-Term Memory (LSTM) network to predict the location of mobile devices using the Time of Arrival (ToA)-based UWB method. Despite all the benefits that come from using supervised models, there are several key challenges ahead. On the one hand, supervised models require labelled LoS/NLoS data, which is both costly and time-consuming, limiting general applicability of supervised models within this context. On the other hand, even minor changes in the indoor environment would require updating the training dataset. Unsupervised learning models~\cite{Fan:2019}, however, eliminate the necessity for labelling of channel conditions, therefore, allowing generalization while saving time and precious resources. Such models, however, are impractical due to the dynamic nature of indoor environments, such as unknown/varying number of users and adverse environmental conditions. Furthermore, energy consumption efficiency is compromised in scenarios where all the beacons require to transmit/receive signals for LoS/NLoS identification.

To tackle the above mentioned issues, the main focus of researchers has been shifted to use Reinforcement Learning (RL) models~\cite{Hajiakhondi:2022, Milioris2019, Mohammadi2018, Li2020} for indoor localization/navigation application. In our previous work~\cite{Hajiakhondi:2022}, for instance, the mobile user is autonomously trained via an RL model to be localized by a set of UWB beacons with LoS connections. The main objective of the RL model within the indoor localization domain is to learn an optimal or near-optimal policy that maximizes the location accuracy. One of the most important challenges of existing RL models~\cite{Hajiakhondi:2022}, however, is that the optimal policy should be learned by the interaction of the agent (mobile user) with the environment (i.e., via trial and error), without any prior information, especially when the model is just initialized. Consequently, it may take a long time for the RL model to reach the optimal policy. Another challenge is the generalization ability of the pre-trained RL model to be used in a new and different environment, where the density/location of obstacles is changing over the time/environment. To tackle  these issues, Uchendu~\textit{et al.}~\cite{Uchendu2022} proposed the Jump-Start RL (JSRL) model, where the agent use a guide-policy instead of a random one at the beginning of the learning process. Consequently,  the learning process is accelerated and the RL generalization ability is highly improved.

\vspace{.025in}
\noindent
\textbf{Contribution:} Motivated by the above discussion, we introduce the Jump-start RL-based Uwb NOde selection ($\JUST$) framework with the application to indoor localization. The main novelty of this work is the design of an autonomous and real-time anchor node selection, where the key objective is to accelerate the location accuracy improvement. Towards this goal, a combination of the guide and exploration policies is used, where the guide-policy significantly speeds up the early learning phase of the RL model to converge to the optimal location accuracy. Furthermore, since any random guide-policy can be used in the $\JUST$ framework, the generalization ability of the pre-trained RL frameworks improves. Simulation results illustrate that the proposed $\JUST$  framework outperforms its state-of-the-art counterparts in terms of the cumulative rewards and location error even in an ultra-dense indoor environment.

The rest of this paper is organized as follows: In Section~\ref{Sec:2}, the system model is provided. Section~\ref{Sec:3} introduces the proposed $\JUST$ framework. Section~\ref{Sec:4} presents experimental results. Finally, Section~\ref{Sec:5} concludes the~paper.

\section{System Model} \label{Sec:2}
In this section, we first introduce the UWB wireless signal model and the TDoA localization formulation. Then, we present the required background on the RL model.

\subsection{UWB Wireless Signal Model} 
In this study, we consider a multi-user indoor environment (e.g., an office  or a hotel building), consisting of $N$ synchronized UWB beacons, denoted by $UWB_i$, $i=1, \ldots, N$, and several mobile users randomly moving through the environment. To support multiple users, we use the Time Hopping (TH) technique as one of the efficient Code Division Multiple Access (CDMA) schemes, where different codes are assigned to distinct users. Given the Pulse Amplitude Modulation (PAM) modulation, the transmitted signal $s_u$ from user $u$ is obtained as
\begin{equation}
s_u(t)= \sum \limits_{n=- \infty}^{\infty} p_u(n) \sum \limits_{s=0}^{N_c-1} c_u(s) w(t-nT_s-sT_c- \theta_u)
\end{equation}
where $T_s$, and $N_c$ represent the symbol time, and the number of chips with duration of $T_c$ in each symbol, respectively. Term $c_u(s)\in \{0,1\}$ denotes the access code associated with the mobile user $u$, and $p_u(n) \in \{-1,1\}$ is the Independent and Identically Distributed (IID) information symbols. Furthermore, term $\theta_u$ is the time asynchronism with uniform distribution $[0,T_s]$, and $w(t)$ is the normalized impulse signal. The received signal $r_i(t)$ by $UWB_i$ is expressed as
%
\begin{equation}\label{e2}
r_i(t) =\sum \limits_{u=1}^{N_u} \sqrt{P_u} \sum_{k=1}^{N(t)} \rho_{u,k}(t,\tau) s_u(t-\tau_{u,k}(t)-\tau_{i})+n(t),
\end{equation}
where $\tau_{u,k}(t)$ and $\rho_{u,k}(t,\tau)= \beta_{u,k}(t,\tau) \exp{(j\Phi_{u,k}(t,\tau))}$  represent the phase delay and the attenuation associated with $k^{\text{th}}$ path, where the total number of detachable paths is denoted by $N(t)$, and $N_u$ is the number of users in the experimental indoor environment. Terms $\beta_{u,k}(t,\tau)$ and $\Phi_{u,k}(t,\tau) $ are the amplitude and phase of $k^{\text{th}}$ path, respectively. Term $\beta_{u,k}(t,\tau)$ is a Nakagami-$m$ random variable that reflects the LoS/NLoS link condition, with $m=1$ representing the Rayleigh fading and $m > 1$ illustrating the Rician channel model. Term  $n(t) \sim \mathcal{N}(0,\sigma ^2)$ is the Additive White Gaussian Noise (AWGN) channel. Moreover, term $\tau_{i}=d_i/c$ is the time of arrival of signal $s_u(t)$, which is equivalent to the delay of the first path within the LoS condition. Finally, terms  $d_i$ and $c=3 \times 10^8$ $m/s$ represent the distance between the user $u$ and $UWB_i$, and the speed of light, respectively. Given the first peak of the estimated Channel Impulse Response (CIR) of at least two UWB beacons $UWB_i$ and $UWB_j$, denoted by $\tau_{i}$ and $\tau_{j}$, the mobile user's location at time slot $t$, denoted by ($x_t, y_t$), is calculated as follows
\begin{eqnarray}\label{e5}
\lefteqn{~~~~~\tau_{i}-\tau_{j}= \nonumber}\\
&&\!\!\!\!\!\!\!\!\!\!\!\!\!\!\!\!\!\!\!\!\frac{\sqrt{\big(x_t-x_i\big)^2+\big(y_t-y_i\big)^2}-\sqrt{\big(x_t-x_j\big)^2+\big(y_t-y_j\big)^2}}{c},
\end{eqnarray}
where ($x_i, y_i$) and ($x_j, y_j$) denote the locations of $UWB_i$ and $UWB_j$, respectively. This completes presentation of the UWB wireless signal model, next, we introduce the RL background.

\subsection{RL Background}
RL model is an area of learning paradigm, where an agent, interacting with the environment, makes a sequence of decisions to achieve the maximum accumulated rewards. Markov Decision Process (MDP) is used to mathematically describe an RL environment, which includes a set of states $\mathcal{S}$, a set of actions $\mathcal{A}$, a reward function $\mathcal{R}$, and a transition function $\mathcal{T}$. Based on the current state $s_t \in \mathcal{S}$ at time slot $t$, the agent takes an action $a_t \in \mathcal{A}$, resulting in a new state $s_{t+1} \in \mathcal{S}$ at time slot $t+1$, which is shown by the transition function $\mathcal{T}(s_t, a_t, s_{t+1})$. The optimum policy $\pi^*$, which leads the agent to the maximum accumulated rewards is given by
\begin{eqnarray}
\pi^* = \arg \max \limits_{\substack{\pi}} \mathbb{E}_{\pi}  \Big \{ \sum \limits_{t=0}^{H-1} \gamma^{t} r_{t+1} | s_0 = s \Big \},
\end{eqnarray}
where $\gamma \in [0,1]$  is the discount factor and $H$ is the total number of steps in one episode. To learn the optimal action, the agent receives a feedback after taking action $a_t$, known as the reward function $r_t=\mathcal{R}(s_t, a_t)$. Q-learning is a value-based RL framework, where the Q-value associated with the action $a_t$ and the state $s_t$ at time slot $t$, denoted by $Q(s_t, a_t)$, is expressed~as
\begin{equation}
Q(s_t, a_t) = \mathbb{E}_{\pi} \Big \{ \sum \limits_{t=0}^{H-1} \gamma^{t} r_{t+1} | s_0 = s, a_0= a, a_t = \pi({s_t}) \Big\}.\label{nEq:33}
\end{equation}
where $\pi$ is the policy that leads to taking an action in a given state. The updated Q-value at each time slot is calculated as
\begin{equation}
Q(s_t, a_t) \leftarrow  (1-\alpha) Q(s_t, a_t)+ \lambda (r_t+ \alpha \max Q(s_{t+1}, a_{t+1})),
\end{equation}
where $\alpha \in [0,1]$ represents the learning rate.

\section{The Proposed $\JUST$ Framework} \label{Sec:3}
\setlength{\textfloatsep}{0pt}
\begin{figure}[t!]
\centerline{\includegraphics [scale = 0.28] {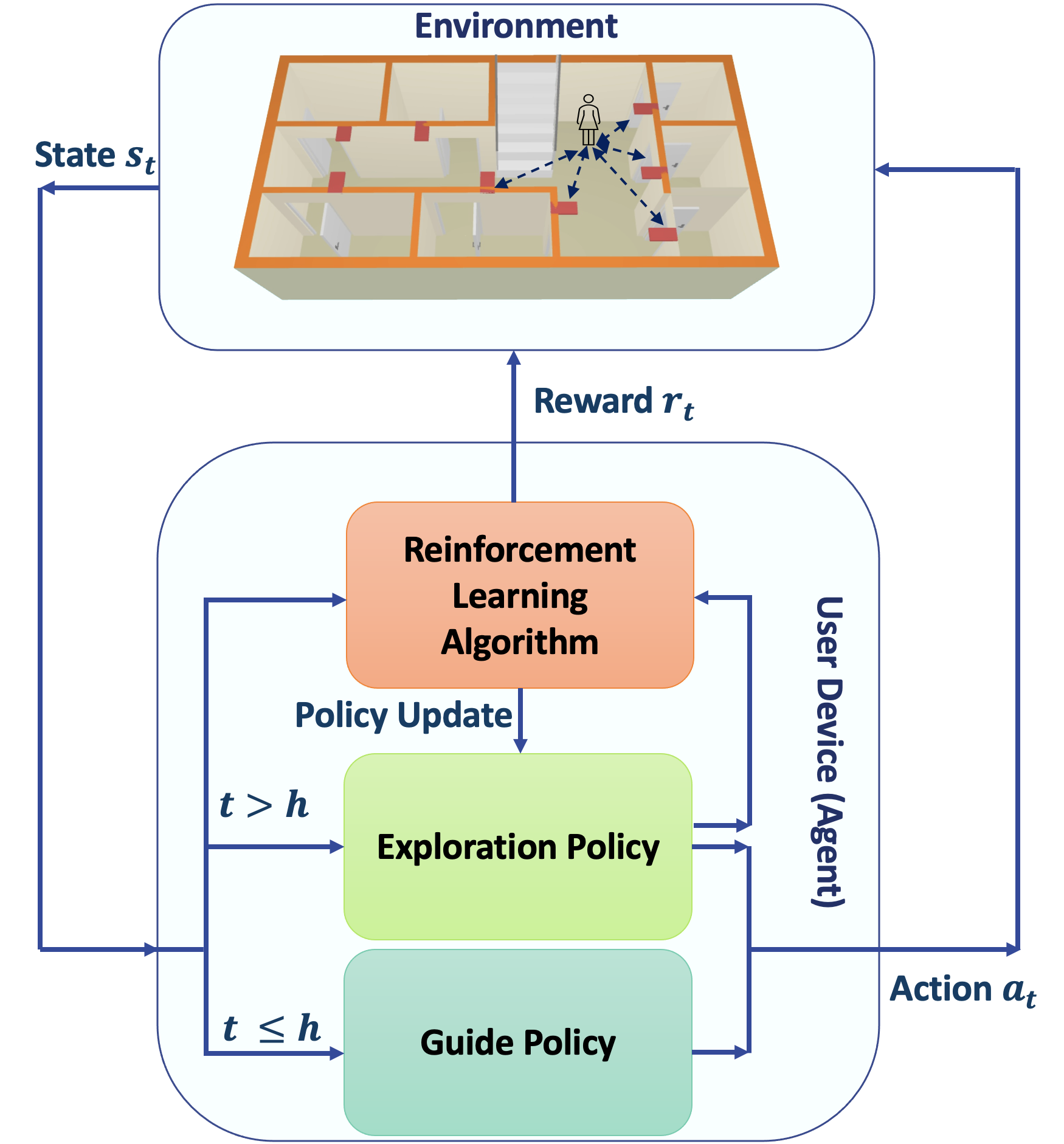}}
\caption{\footnotesize The block diagram of the proposed $\JUST$ anchor selection framework.}
\label{sys}
\end{figure}
In this section, we first introduce the JSRL model, and then present details of the proposed $\JUST$ framework.

\subsection{JSRL model}
Conventionally, an RL-based agent selects a random action $a_t$ at the beginning of the learning process, where there is no prior information resulting in a long time to reach the optimal policy $\pi$. The main difference between the conventional RL and JSRL model~\cite{Uchendu2022} is that the agent in the JSRL framework has access to two policies, called guide-policy  $\pi^g(a|s)$, and the exploration policy  $\pi^e(a|s)$. While the exploration policy is the same policy used in conventional RL models, which will be updated during the training, the guide-policy is a fixed prior policy, learned via an RL model or manually/randomly constructed. Since the Q-table is initialized with zeros, the agent of the JSRL model follows  $\pi^g(a|s)$ instead of randomly selecting an action $a_t$ based on the untrained policy $\pi^e(a|s)$. Considering the fact that the distribution of the data under policy   $\pi^g(a|s)$ is not exactly the same as the policy $\pi^e(a|s)$, the state space of the policy $\pi^e(a|s)$ would be different. Therefore, it is essential to gradually transit the data collection from the  policy $\pi^g(a|s)$ to  policy $\pi^e(a|s)$. Consequently, given an RL model with horizon $H$, we define guide-step $h \leq H$ as the number of steps where the agent uses policy $\pi^g(a|s)$, initialized with $H$ and gradually decreases over the course of training. More precisely, the action is selected based on  $\pi^g(a|s)$ for $h$ steps at the initial stage of each training episode, continuing with $\pi^e(a|s)$ for remaining $(H-h)$ steps.

\subsection{$\JUST$ Anchor Selection}
Following Reference~\cite{Uchendu2022}, we use the JSRL model in our proposed $\JUST$  framework to accelerate the learning process. Due to the dynamic nature of indoor venues, caused by varying environmental conditions, the proposed $\JUST$ framework seeks to train the mobile user to autonomously find the optimal LoS connections at any given time and place. The proposed $\JUST$  framework consists of the following main components:

\vspace{.1in}
\noindent
\textbf{(i) Agent:} In the proposed $\JUST$ framework, the mobile user operates as the agent, interacting with the environment via a set of actions.

\vspace{.1in}
\noindent
\textbf{(ii) State-Space:} The state $\mathbf{s}_t \in \mathcal{S}$ is defined as the user's location $(x_t,y_t)$ at time slot $t$. Following Reference~\cite{Li2020}, we discretize the indoor environment into $N_l= N_x \times N_y$ points, where $x_t$ and $y_t$ are obtained as
\begin{eqnarray}\label{eq1}
x_t&=& x_{t-1} + \zeta_x\label{e7},~~~~~0 \leq x_t \leq N_x\\
\text{and } y_t&=& y_{t-1} + \zeta_y,~~~~~0 \leq y_t \leq N_y \label{eq2}.
\end{eqnarray}
where $ \zeta_x, \zeta_y \in \{-1,0,1\}$ are random numbers, indicating the user's movement in $x$-axis and $y$-axis~\cite{Li2020}, respectively. It should be noted that although RL models with a continuous and high-dimensional state-space~\cite{Marchesini2020} provide higher resolution and precise localization, they suffer from high complexity, making them  inefficient for real-time applications requiring low latencies. The benefits of the discrete RL models in the context of indoor localization~\cite{Li2020, Mohammadi2018}, therefore, make us to choose discrete RL over its continuous counterpart.

\vspace{.1in}
\noindent
\textbf{(iii) Action-Space:} The action space is defined as a set of nearby UWB beacons, where the user's location can be determined by extracting the time information from the received signal of the corresponding beacons. The cardinality of the action space, denote by $N_a$, is given by

\begin{equation}
N_a= \frac{N_u!}{(N_u-N_r)! \enspace N_r!},
\end{equation}
where $N_r$ represents the number of required beacons for localization, depending on some parameters, such as the indoor localization framework (i.e., ToA, TDoA, and Two Way Ranging (TWR)), and the dimension of the experimental environment, i.e., 2D or 3D area. Considering the fact that at least two UWB beacons are required for the TDoA-based localization scheme in a 2D environment, the selected action is a vector, denoted by $\mathbf{a}=[a_i,a_j]$, where $a_i$, $a_j$ represent $UWB_i$ and $UWB_j$, respectively.

\vspace{.1in}
\noindent
\textbf{(iv) Reward:} As stated previously, the main objective of the proposed $\JUST$ framework is to minimize the location error caused by UWB beacons with NLoS connections. Therefore, after taking action $\mathbf{a}_t$, the estimated user's location $(x_{es,t}^{(i,j)},y_{es,t}^{(i,j)})$ is calculated, where the superscript $(i,j)$ indicates that the estimated location is obtained by the received signals of  $UWB_i$ and $UWB_j$. Considering the fact that even if one of these two connections is NLoS, we will face with a remarkable location error, the combination of $UWB_i$ and $UWB_j$ at any location/time is of paramount importance. For this reason, the reward function $\mathcal{R}(s_t, a_t)$ is defined as
\begin{eqnarray} \label{e18}
\!\!\!\!\!\!\!\!\!\!\mathcal{R}(s_t, a_t)&\!\!\!\!=\!\!\!\!\!&\left\{\begin{array}{ll}
\dfrac{1}{\mathcal{E}_t} ,~~~~~~~~\mbox{$\mathcal{E}_{t} \leq \mathcal{E}_{th},$}\\
- \mathcal{E}_t,~~~~~~ \mbox{o.w.}\end{array}\right. ,
\end{eqnarray}
where $\mathcal{E}_{th}$ is a pre-defined threshold value for the maximum acceptable location error~\cite{Mohammadi2018}, and $\mathcal{E}_t$ denotes the location error at time slot $t$, calculated as
\begin{equation}\label{e19}
\mathcal{E}_t=\sqrt{(x_t-x_{es,t}^{(i,j)})^2+(y_t-y_{es,t}^{(i,j)})^2}.
\end{equation}
This completes presentation of the proposed JUMP anchor selection framework, next, we will describe our testbed and simulation results.

\begin{figure*}[h]
\centering
\mbox{\subfigure[]{\includegraphics[scale = .26]{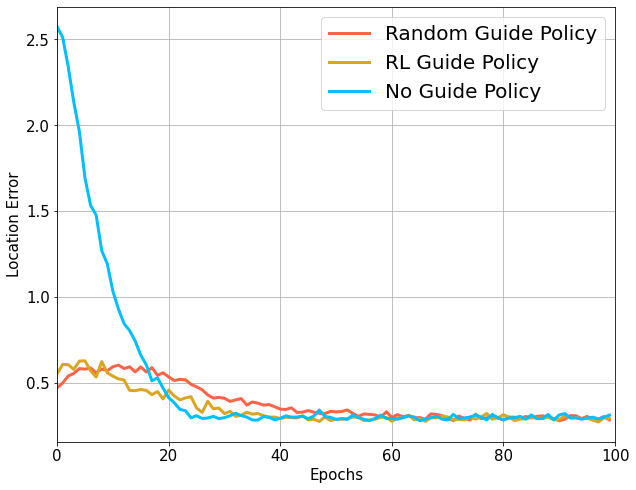}}
\subfigure[]{\includegraphics[scale = .22]{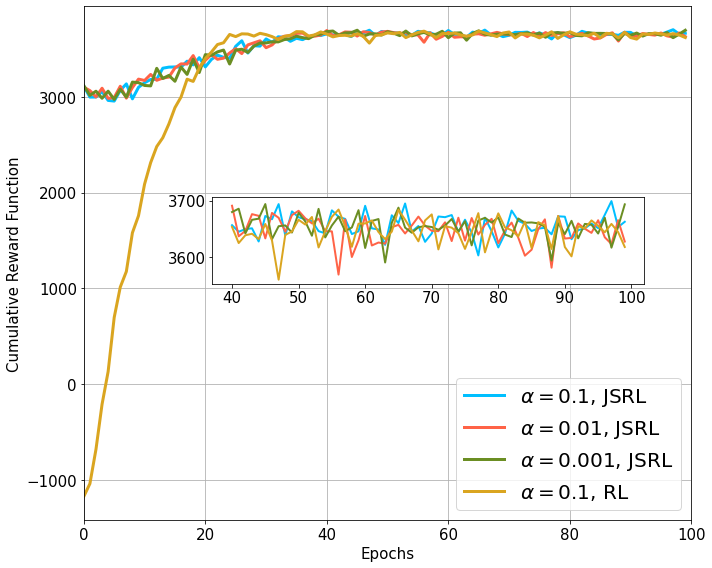}}
\subfigure[]{\includegraphics[scale = .227]{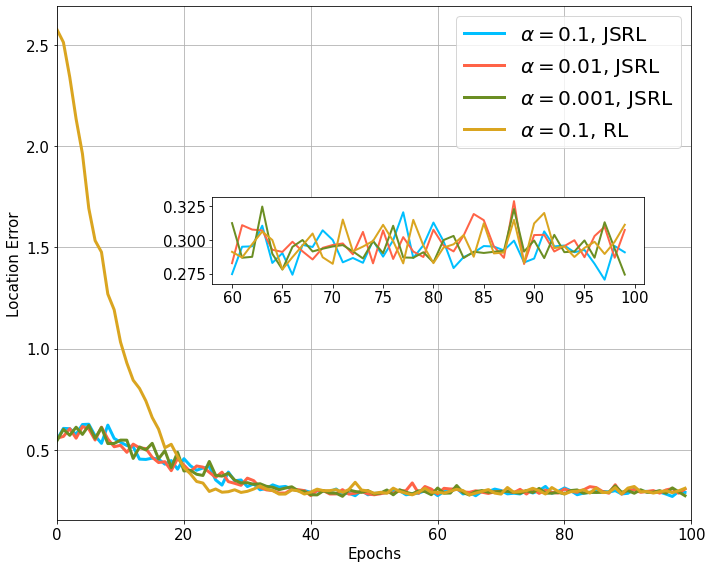}}}
\vspace{-.15in}
\caption{\footnotesize Investigating the effect of (a) Guide-policy on the location error; (b) Learning rate $\alpha$ on the cumulative rewards, and; (c) Learning rate $\alpha$ on the location error.}\label{Fig:10}
\end{figure*}

\begin{figure*}[t!]
\centering
\mbox{\subfigure[]{\includegraphics[scale = .22]{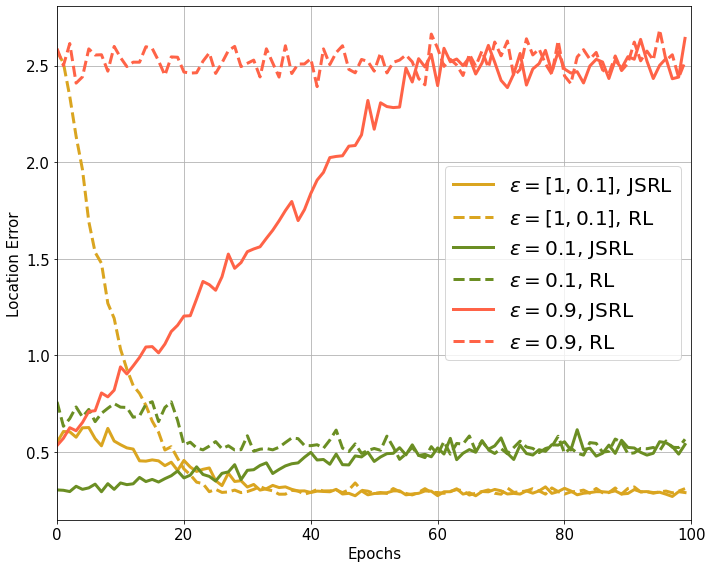}}
\subfigure[]{\includegraphics[scale = .26]{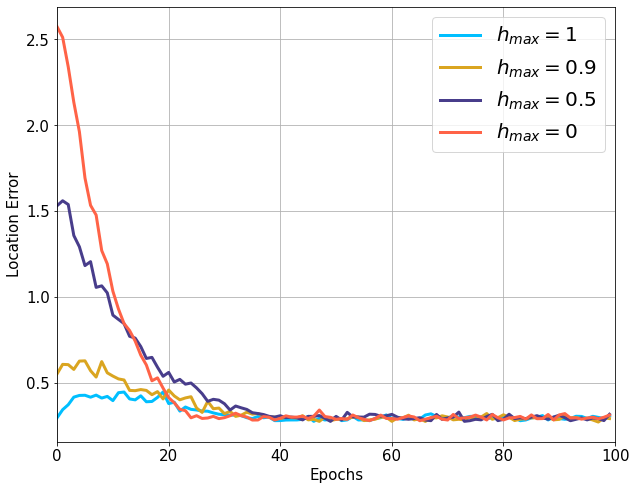}}
\subfigure[]{\includegraphics[scale = .26]{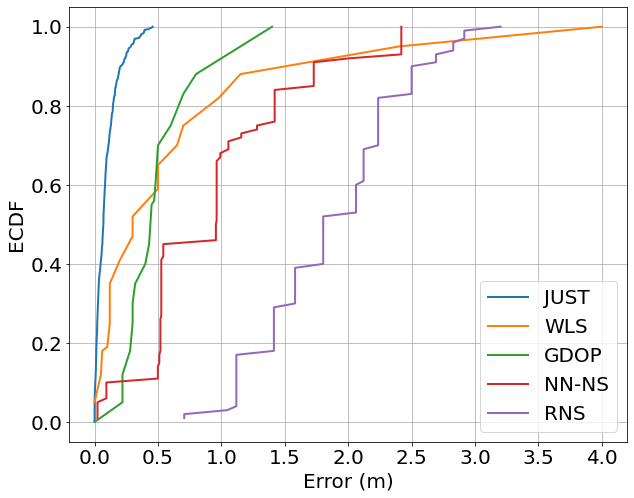}}}
\vspace{-.1in}
\caption{\footnotesize Investigating the effect of (a) $\epsilon$ on the location error; (b) $h_{max}$  on the location error, and; (c) ECDF on the location error.}\label{Fig:2}
\end{figure*}

\begin{figure*}[h]
\centering
\mbox{\subfigure[]{\includegraphics[scale = .265]{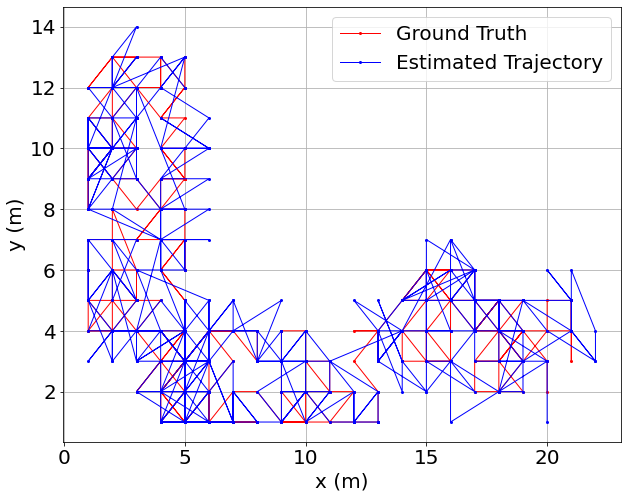}}
\subfigure[]{\includegraphics[scale = .265]{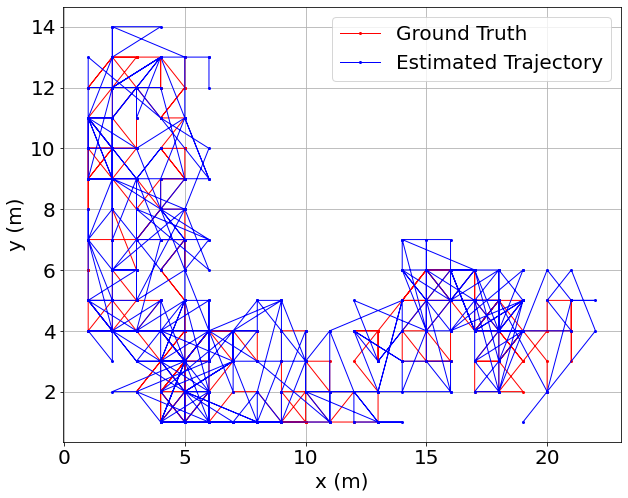}}
\subfigure[]{\includegraphics[scale = .265]{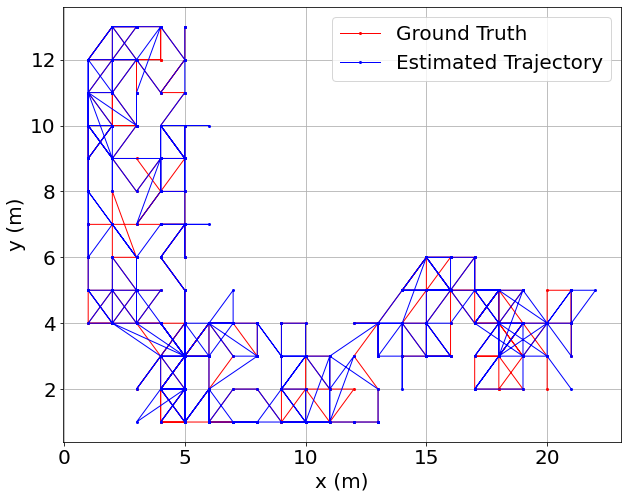}}}
\vspace{-.1in}
\caption{\footnotesize Ground truth and estimated random trajectories by using: (a)  Nearest neighbor; (b) Random, and; (c) $\JUST$ frameworks.}\label{Fig:12}
\end{figure*}

\section{Simulation Results}\label{Sec:4}

To evaluate the performance of the proposed $\JUST$ framework, we consider an experimental indoor area such as an office building with the size of ($60 \times 50$) $m^2 $, which is compromised of several non-overlapping sub-areas~\cite{Wang:2016, Li2020}. Following Reference~\cite{Li2020}, each sub-area is discretized into several square zones, where the dimension of each zone is ($1 \times 1$) $m^2$. Although the location resolution, i.e., the number of discretized points in the indoor environment $N_l$, is proportional to the location accuracy, it also results in higher state-space, complexity, and the respond-time. Therefore, there should be a trade-off between the location resolution and the respond-time of the learning model. Despite the recent RL-based localization works~\cite{Li2020, Mohammadi2018}, where the environment is divided into a grid of $(5 \times 5)$ $m^2$ and $(3 \times 3)$ $m^2$ cells, respectively, we assume higher resolution of $(1 \times 1)$ $m^2$ to improve the location accuracy. Mobile users are randomly moving through the network in $8$ directions based on Eqs.~\eqref{eq1} and~\eqref{eq2}, where it is assumed that mobile users are placed at zones' center~\cite{Li2020}. At each location, the transmitted signal by the mobile user are received by a set of nearby UWB beacons. Due to the obstacles in the environment, the received signal would be LoS or NLoS connections, where it is assumed that the channel condition of $UWB_i$, for ($1\leq i \leq N_u$), is determined randomly at each zone to initialize the environment.

Fig.~\ref{Fig:10}(a) illustrates the effect of the guide-policy on the proposed $\JUST$ framework. As shown in Fig.~\ref{Fig:10}(a), using a random Q-table or the one obtained by an RL model as the guide-policy outperforms the conventional RL approach, accelerating the learning process of the $\JUST$ framework, improving the location accuracy, and increasing the RL's generalization capabilities. We also investigate the effect of learning rate $\alpha$ on the proposed $\JUST$ framework to obtain the best value of $\alpha$. As shown in Figs.~\ref{Fig:10}(b)-(c), increasing the number of epochs decreases the location error and increases the cumulative rewards, illustrating that the model is well-trained. Moreover, it is evident that the learning rate has not a great impact on the location accuracy.

Fig.~\ref{Fig:2}(a) shows the effect of $\epsilon$ on the $\JUST$ framework, as a parameter to maintain a trade-off between exploration and exploitation, where the random action $a_t$ is chosen with the probability of $\epsilon$. Note that $\epsilon = [1, 0.1]$ means that $\epsilon$ is initialized with $1$, gradually decreasing with time by $\Delta \epsilon=\dfrac{\epsilon_{max}-\epsilon_{min}}{N_{epoch}} $ to $0.1$, with $N_{epoch} = 100$, which is the best strategy according to the results of Fig.~\ref{Fig:2}(a). Moreover, Fig.~\ref{Fig:2}(b) illustrates the maximum step $h_{max}$ that the guide-policy is initially used, where $h_{max}=0$ represents the conventional RL model with no guide-policy. As shown in Fig.~\ref{Fig:2}(b), larger $h_{max}$ results in the lower location error.

To illustrate the effectiveness of the proposed framework, we compare it with four baseline models: (i)  Weighted Least Square (WLS) anchor selection~\cite{Albaidhani:2019}; (ii) Geometric Dilution of Precision (GDOP) anchor selection~\cite{Albaidhani:2020}; (iii) Nearest Neighbor Node Selection (NN-NS), where the mobile user is localized by $N_r$ number of nearest beacons, and; (iv) Random Node Selection (RNS), where a set of UWB beacons are randomly selected for localization. Fig.~\ref{Fig:2}(c) compares the Empirical Cumulative Distribution Function (ECDF) of different frameworks. According to the results shown in Fig.~\ref{Fig:2}(c), the location error caused by the proposed $\JUST$ framework is considerably lower than that of its counterparts. Finally, in Fig.~\ref{Fig:12} we consider a random trajectory in a $(24 \times 15)$ $m^2$ rectangular indoor environment, and compare the ground truth with the estimated trajectories by our proposed $\JUST$ framework and two other baselines. As shown in Fig.~\ref{Fig:12}, the estimated path by our proposed framework in the most points closely follows that of the ground~truth.

Finally, we use the Root Mean Squared Error (RMSE), which is a generally used performance metric within the localization domain, calculated as follows
\begin{equation}
RMSE= \sqrt{\dfrac{\sum \limits_{t=1}^{N} (L_t - L_{est,t})^2}{N}},
\end{equation}
where $L_t =(x_t,y_t)$ and $ L_{est,t}= (x_{es,t}, y_{es,t}) $ represent the exact and the estimated location of the user at time $t$, and $N$ denotes the total number of steps once the proposed $\JUST$ framework reaches the steady state. The proposed $\JUST$ framework achieves the localization error of $0.32$ m, while the CNN~\cite{Nguyen:2020} and LSTM-based~\cite{Poulose:2020} localization frameworks track a mobile user with $0.58$, $0.48$ location errors, respectively.

\section{Conclusion}\label{Sec:5}
In this paper, we presented the Jump-start RL-based Uwb NOde selection ($\JUST$) framework with  application to indoor localization. The key objective was to overcome several challenges in existing UWB-based anchor selection frameworks, such as not adapting with time-varying environmental conditions, lack of generalization, and taking long times to reach the optimal policy. Using the proposed $\JUST$ framework, the learning process is accelerated, where the mobile user is autonomously trained to identify a set of UWB beacons with LoS connections to be localized based on the 2-D TDoA technique. The effectiveness of the proposed framework is evaluated in terms of the location error and cumulative rewards. Simulation results illustrated that the proposed $\JUST$ framework reduced the location error. According to the simulation results, the proposed framework outperformed its counterparts in tracking a mobile user moving in a random trajectory. With the emphasis on non-uniform distribution of UWB beacons, our future research direction is to propose an adaptive version of the $\JUST$ framework, where the cardinality of the action space can be  adjusted  during the learning process.


\end{document}